# Visual Brightness Characteristics of Starlink Generation 1 Satellites

Anthony Mallama and Jay Respler

2022 October 30

Contact: anthony.mallama@gmail.com

Abstract

A large dataset of visual magnitudes for all three designs of Starlink satellites is analyzed. Brightness phase functions are derived for the Original, VisorSat and Post-VisorSat models. Similarities and differences between the functions for these spacecraft are noted. A metric called the characteristic magnitude is defined as the average brightness of a satellite when seen overhead at the end of astronomical twilight. When the phase functions are evaluated according to this metric, the characteristic magnitudes are: Original, 4.7; VisorSat, 6.2; and Post-VisorSat, 5.5.

Introduction

Starlink satellites comprise the largest constellation of spacecraft currently in orbit. There are presently about 3,000 and many thousands more are planned. These objects are already interfering with the work of researchers making astronomical observations (Mroz et al., 2022) and they are a distraction for amateur astronomers (Mallama and Young, 2021). The International Astronomical Union has established a Centre for the Protection of Dark and Quiet Skies from Satellite Constellation Interference to address this problem.

SpaceX has launched three different models of Generation 1 Starlink satellites so far. These will be followed in coming years by Generation 2 satellites which are expected to be larger and possibly brighter. This paper addresses the visual brightness of the Generation 1 satellites. The first spacecrafts, designated as Original here, are very bright and they generated immediate concern from astronomers. SpaceX addressed this problem by adding a sun shade to the second model which is called VisorSat. The shade reduced the amount of sunlight reflecting off the satellites to observers on the ground and made them less bright. However, SpaceX stopped installing sun shades when they added laser communication capability to the satellites because the visors blocked the beams. This third model is termed Post-VisorSat here.

This paper characterizes the visual brightness of all three models of Generation 1 Starlink satellites. Section 2 describes the observations and the data processing. Section 3 addresses the phase functions that quantify the brightness of the three models. Section 4 presents the conclusions. A reading list of papers on satellite constellations and their impact is provided in Appendix A.

2. Observations and data processing

The authors have recorded apparent visual magnitudes of Starlink satellites since they were first launched in 2019. Our geographic coordinates are $40.330^o$ N, $74.445^o$ W for Respler and $38.982^o$ N, $76.763^o$ W for Mallama. Most of the observations were made with binoculars though a few of the brighter ones were acquired by eye alone and some of the fainter ones were acquired through telescopes. Satellite magnitudes are measured by comparing their brightness to reference stars in the same field of view. The method is described in more detail by Mallama (2022a). The observations analyzed in this paper are for satellites at their operational altitudes which are around 550 km.

In order to characterize brightness, the apparent magnitudes are adjusted from the actual range to a standard distance of 1,000 km by applying a factor that corresponds to the inverse square law of light. The parameter of characterization, called the phase angle, is that arc measured at the satellite between the directions to the Sun and to the observer. This phase angle is also computed in this data processing step of the analysis.

Finally, the distance-adjusted magnitudes are fitted by least squares to the phase angles. The resulting quadratic polynomial fit, called the phase function, is the method of characterization used in this paper.

3. Brightness characterization

The phase functions for the Generation 1 models of Starlink spacecrafts are illustrated separately and comparatively in Figure 1. The corresponding quadratic coefficients for the three models are listed in Table 1.

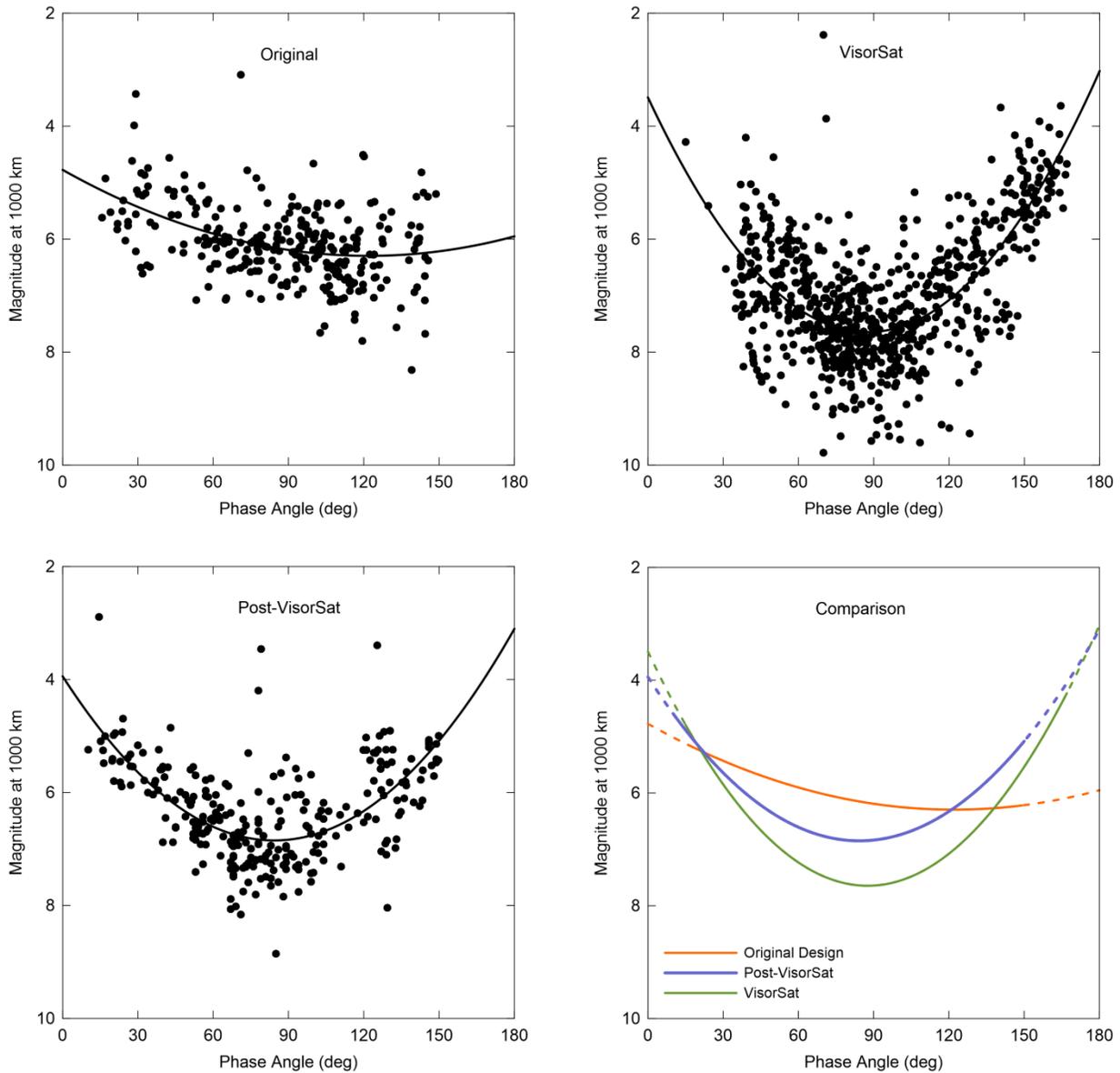

Figure 1. The observations and best fit functions for the three models of Generation 1 Starlink satellites are illustrated in the upper left, upper right and lower left panels. The three phase curves are compared in the lower right panel. Dotted lines are extrapolations beyond the observed limits.

Table 1. Polynomial coefficients for the phase functions

| Order | 0 | 1 | 2 |
|---|---|---|---|
| Original | 4.774 | 0.02496 | -0.0001023 |
| VisorSat | 3.493 | 0.09481 | -0.0005412 |
| Post-VisorSat | 3.944 | 0.06893 | -0.0004089 |

The observations plotted in Figure 1 reveal a considerable amount of scatter. This dispersion is due mostly to the complex pattern of sunlight reflecting from the satellites. Nevertheless, distinctly different phase functions are apparent for the three models of Starlink satellites.

The Original spacecrafts have a relatively flat phase function, so they are comparatively bright over a wide range of phase angle. VisorSats have a deep U-shaped phase function, so they are fainter than the Originals at phase angles around 90$^o$ but they are brighter at angles closer to 0$^o$ and 180$^o$. As seen from the ground, this means that VisorSats tend to be fainter than Original satellites when near zenith and brighter when they are low in the sky and their azimuth is toward or away from the Sun. Finally, the phase function for Post-VisorSat spacecraft is intermediate between those of Original and VisorSat.

A useful metric for visual observers, called the *characteristic magnitude,* is the average apparent brightness of a satellite when seen overhead at the end of astronomical twilight. The phase angle corresponding to this geometry is 72$^o$, as illustrated in Figure 2, and the distance is adjusted to 550 km for Starlink Generation 1 satellites because that is their approximate altitude. When the phase functions are evaluated for this angle and distance, the characteristic magnitudes are: 4.7 (Original), 6.2 (VisorSat) and 5.5 (Post-VisorSat).

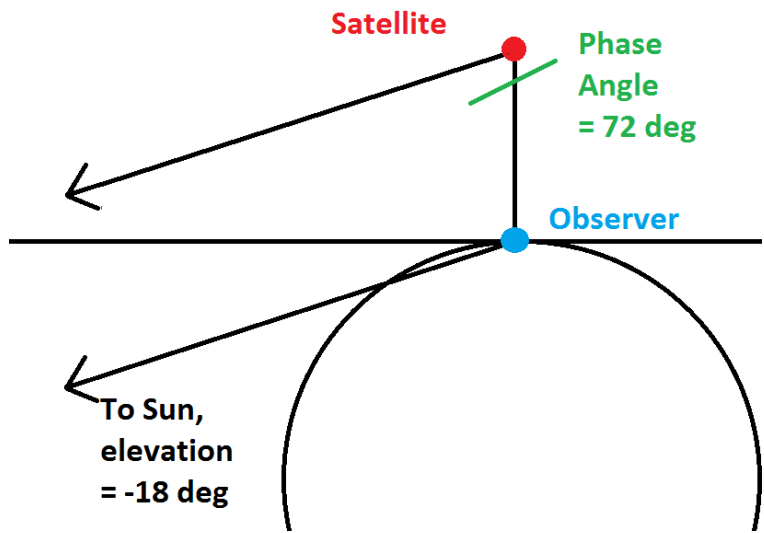

*Figure 2. The geometry associated with the characteristic magnitude. The Sun is 18° below the horizon at the end of astronomical twilight, so the phase angle is 72°. Satellites interfere strongly with astronomical observations at this time because many of them are still illuminated by the Sun.*

4. Conclusions

A large set of apparent visual magnitudes for all three designs of Generation 1 Starlink satellites are characterized. Phase functions are derived for these designs and differences are noted. Characteristic magnitudes are computed for the three models.

Appendix A. Reading list of papers on satellite constellation interference

These papers are grouped by subject matter as follows: impacts on observations, working group reports and conference material, satellite brightness, mitigation strategies and general. Titles are given here to indicate content while full citations are provided in the Reference section.

Appendix A-1. Impacts on celestial observations

Gallozzi et al. 2020. Concerns about ground based astronomical observations: a step to safeguard the astronomical sky.

Hainaut and Williams, 2020. Impact of satellite constellations on astronomical observations with ESO telescopes in the visible and infrared domains.

Appendix A-2. Working group reports and conference proceedings

Appendix A-3. Satellite brightness

Appendix A-4. Mitigation strategies

Appendix A-5. General audience

Mallama and Young, 2021. The satellite saga continues.

Witze, 2022. 'Unsustainable': how satellite swarms pose a rising threat to astronomy.